\documentclass[aps,prl,amsmath,amssymb,twocolumn,showpacs]{revtex4}
\usepackage{amssymb}
\usepackage{graphicx}
\begin{document}

\title{Evidence of anomalous dispersion of the generalized sound velocity in glasses}

\author{
B.~Ruzicka$^1$, T.~Scopigno$^1$, S.~Caponi$^2$, A.~Fontana$^3$,
O.~Pilla$^3$, \\P.~Giura$^4$, G.~Monaco$^4$, E.~Pontecorvo$^1$,
G.~Ruocco$^1$, F.~Sette$^4$
        }
\affiliation{
$^1$ INFM and Dipartimento di Fisica, Universit\'a di Roma "La Sapienza", 00185 Roma, Italy \\
$^2$  INFM and Dipartimento diFisica, Universit\'a di L'Aquila, 67100, L'Aquila, Italy \\
$^3$ INFM and Dipartimento di Fisica, Universit\'a di Trento,38050, Trento, Italy \\
$^4$ European Synchrotron Radiation Facility, BP 220, 38043
Grenoble, France}
\date{\today}
\begin{abstract}
The dynamic structure factor, $S(Q,\omega)$, of {\it vitreous
silica}, has been measured by inelastic X-ray scattering in the
exchanged wavevector ($Q$) region $Q$=4$\div$16.5 nm$^{-1}$ and up
to energies $\hbar\omega$=115 meV in the Stokes side. The
unprecedented statistical accuracy in such an extended energy
range allows to accurately determine the longitudinal current
spectra, and the energies of the vibrational excitations. The
simultaneous observation of two excitations in the acoustic
region, and the persistence of propagating sound waves up to $Q$
values comparable with the (pseudo-)Brillouin zone edge, allow to
observe a positive dispersion in the generalized sound velocity
that, around $Q$$\approx$5 nm$^{-1}$, varies from $\approx$6500 to
$\approx$9000 m/s: this phenomenon was never experimentally
observed in a glass.

\end{abstract}
 \maketitle

The stimulating evidence of the existence of short wavelength
phonon-like excitations sustained by non periodic solids, as
demonstrated by recent Inelastic X-ray Scattering (IXS) studies
\cite{IXSglasses}, has motivated a massive number of
investigations. This notwithstanding, a general consensus on the
{\it characteristics} of observed excitations is still lacking
\cite{Consensus}. Among several specific questions waiting for an
answer, one of the most important regards the existence and the
features of the {\it dispersion relations} in the whole first
pseudo-Brillouin Zone, and their connection with the well known
universal anomalies of glasses (low temperature transport and
thermodynamic properties \cite{Anomalies}, excess in the density
of states \cite{BP}, etc.).

In the specific case of vitreous silica, the archetype of the
strong glasses, up to now the experimental studies were limited to
i) the low $Q$ region ($Q$$<$4 nm$^{-1}$) -investigated by IXS-
where the spectra are described by a single excitation with energy
position ($\hbar\Omega(Q)$) linearly dispersing with $Q$
\cite{IXSsilica1,IXSsilica2,IXSsilica3}, and ii) the high $Q$
region ($Q$$\geq$15 nm$^{-1}$) where Inelastic Neutron Scattering
(INS) data (confined to the high $Q$ and low $\omega$ portion of
the $S(Q,\omega)$ due to the kinematic restrictions) show again
spectra dominated by a single excitation, but having now a $Q$
independent energy \cite{BP,INSsilica}. This scenario led to the
conclusion that the longitudinal acoustic branch in vitreous
silica -dispersing with a sound velocity $v$$\approx$6400 m/s at
low $Q$- flattens at increasing $Q$ giving origin to the Boson
Peak \cite{Courtens_LA=BP}. However, as the cross-over between the
low- ($\Omega(Q)$=$vQ$) and high-$Q$ ($\Omega(Q)$=$\Omega_{BP}$)
regimes lies in the not-explored $Q$ region, it was not possible
to establish a firm conclusion on the shape of the dispersion
relation of this mode. A hint on this issue came from Molecular
Dynamic (MD) simulations. Surprisingly, the MD works showed the
simultaneous presence of two excitations in the longitudinal
current spectra at large enough $Q$ ($Q$$>$5 nm$^{-1}$)
\cite{MDdellanna,MDkob,MDelliott,MDpilla}. MD indicated that the
excitation detected at low $Q$ by IXS corresponds to the
Longitudinal Acoustic (LA) mode, while that detected at high $Q$
by INS is associated to the spilling of the Transverse Acoustic
(TA) mode in the longitudinal spectra and not to the longitudinal
branch \cite{MDpilla}. The presence of the signature of the
transverse dynamics in the {\it longitudinal} spectra is not
surprising. Indeed, in a disordered medium, the polarization
character of the modes becomes ill defined at short wavelengths
-when the sound waves no longer see the system as a continuum- and
this gives rise to modes with mixed polarization \cite{Mixing}.

The MD works \cite{MDkob,MDelliott,MDpilla} also indicated the
presence, in vitreous silica, of another important feature of the
LA branch: a positive dispersion of the generalized sound
velocity, $v(Q)$=$\Omega(Q)/Q$. A positive dispersion of $v(Q)$ is
usually observed in viscoelastic liquids where a step-like change
of $v(Q)$ is found when the condition
$\Omega(Q)\tau_\alpha(Q,T)$=1 is fulfilled, being
$\tau_\alpha(Q,T)$  the structural ($\alpha$) relaxation time.
This condition marks the transition from the low frequency (and
low $v$) viscous behavior to the high frequency (and high $v$)
elastic behavior of the liquid. In glasses, being
$\tau_\alpha(Q,T)$ basically {\it infinite}, the MD observed
positive dispersion cannot be associated to the structural
relaxation process but to a further relaxation process named {\it
instantaneous} or {\it microscopic}. This process, already
observed in the MD simulation of monatomic glasses \cite{relHarm}
and, experimentally, in liquid metals above melting \cite{LiqMet},
has been so far explained in term of {\it structural disorder}
\cite{relHarm}. No experimental evidences for such a positive
dispersion in glasses have been reported yet.

In this Letter we present an IXS study of the dynamic structure
factor of vitreous silica in the intermediate $Q$ region
(4$<$$Q$$<$16.5 nm$^{-1}$). Using also existing IXS data at
$Q$$<$4 nm$^{-1}$, we draw a self-consistent picture of the whole
pattern of excitations in the first pseudo-Brillouin Zone (p-BZ)
of this prototype glass. Specifically, we find that i) the LA
branch spans energies as high as 80 meV, confirming and extending
the validity of previous IXS studies
\cite{IXSsilica1,IXSsilica2,IXSsilica3} limited to $Q$$<$4
nm$^{-1}$ and to $E$$<$20 meV; ii) above $Q$=4 nm$^{-1}$ the
spectra show the simultaneous presence of {\it two} excitations,
the high energy one associated to the LA modes, and the low energy
one assigned to the TA excitation. The latter excitation appears
in the longitudinal spectra due to the mixing phenomenon
\cite{Mixing}, and its experimental observation confirms previous
MD results \cite{MDdellanna,MDkob,MDelliott,MDpilla} and parallels
similar findings in glassy glycerol \cite{IXSgly}. iii) Finally,
and most importantly the clear identification of longitudinal
acoustic branch allows to observe experimentally the existence in
glasses of a positive dispersion of the generalized longitudinal
sound velocity. The transition is found at $Q$$\approx$5
 nm$^{-1}$, where $v$ undergoes a change from $\approx$6400 to
$\approx$9000 m/s).

The measurements were performed in two different sessions at the
inelastic X ray scattering beamline ID16 of the European
Synchrotron Radiation Facility (ESRF). The experimental setup
consists of a backscattering operating monochromator and of a 6.5
m analyzer arm hosting a five analyzers bench with a constant
angular offset between each analyzer. The incident energy was
17794 eV, corresponding to the Si(9,9,9) reflection. In this
configuration the overall (FWHM) energy and $Q$ resolutions were
$\delta E =3.0$ meV and $\delta Q=0.27$ nm$^{-1}$, respectively.
The exchanged wavevector is selected by rotating the analyzer arm.
Each energy scan was performed at constant momentum transfer by
varying the relative temperatures of the monochromator and the
analyzer crystals. Further details on the beamline are reported
elsewhere \cite{Beamline}. The sample of vitreous silica
(SUPRASIL), heated at T=1270 K, was studied in a Q range not
covered by previous measurements, i.~e. at $Q$=4, 5, 6 in a first
run and from 6.5 to 16.5 nm$^{-1}$ in step of 2.5 nm$^{-1}$ in a
second run. In order to follow the evolution of the inelastic
signal, in the second run we performed wide energy scans up to 115
meV. To achieve the necessary statistical accuracy the spectra
were collected on the Stokes side only (and on a small portion of
the anti-Stokes to cover most of the central peak): the total
integration time was 30 minutes/point, i.~e. $\approx$120 hours
for the collection of a whole spectrum.

\begin{figure}[h]
\centering
\includegraphics[width=.45\textwidth]{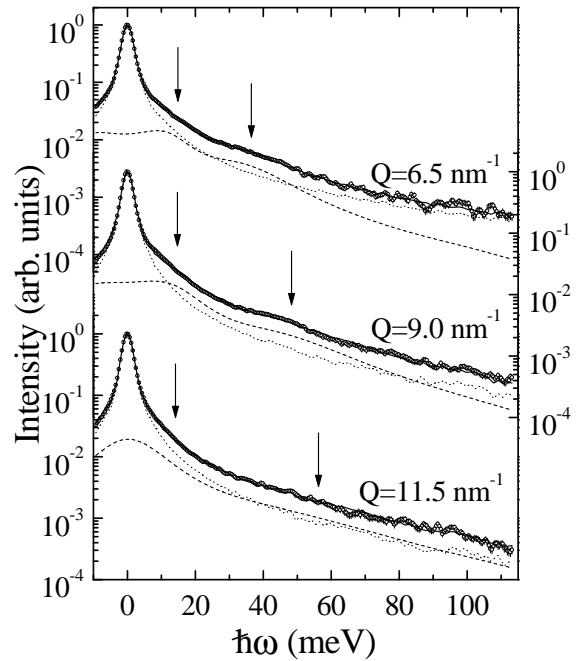}
\vspace{-2.1cm} \caption{IXS spectra of vitreous silica at
$T$=1270 K for selected values of the exchanged wavevector in the
region of interest (open circles with error bars). The total
fitting result with two excitations (full line) is also reported,
along with the elastic (dotted line) and inelastic (dashed lines)
contributions. The arrows indicate the approximate energy
location of the excitations "bumps".} \label{fig1}
\end{figure}

As an example, the spectra obtained for three selected $Q$ values
are shown in Fig.~1. The IXS signal (open circles), proportional
to the dynamic structure factor, $S(Q,\omega)$, convoluted with
the experimentally determined instrumental resolution function,
$R(\omega)$, is reported in log scale together with the elastic
(dotted line, coincident with $R(\omega)$) and inelastic (dashed
line) contributions as derived from the fit (see below). Directly
from the IXS data -at least for the lowest two reported $Q$
values- one can infer the presence of two shoulders, which are not
present in the smoothly decreasing resolution function. These
shoulders, indicated in the figure by arrows, point out the
presence of two excitations. To get quantitative information on
the energies of the excitations and on their $Q$-dependences, the
data have been fitted by the convolution of $R(\omega)$ with a
model function for $S(Q,\omega)$ weighted by the detailed balance
function. The model function is the sum of an elastic and an
inelastic contribution,
$S(Q,\omega)=S_{el}(Q,\omega)+S_{inel}(Q,\omega)$; the former,
accounting for the frozen structural relaxation, is represented by
a delta function, the latter has been tentatively described by one
or two excitation(s) model through, respectively, one or two
Damped Harmonic Oscillator function(s). This model function has
been chosen only to extract information on the position of the
peaks. Indeed, a full description of the spectral shape in the
investigated $Q$ region would need a detailed generalized
hydrodynamic model, where the transverse and the longitudinal
variables couple each other due to the broken ergodicity in the
glass. The use of this detailed model to describe the spectra goes
beyond the aim of the present Letter.

\begin{figure}[h]
\centering
\includegraphics[width=.43\textwidth]{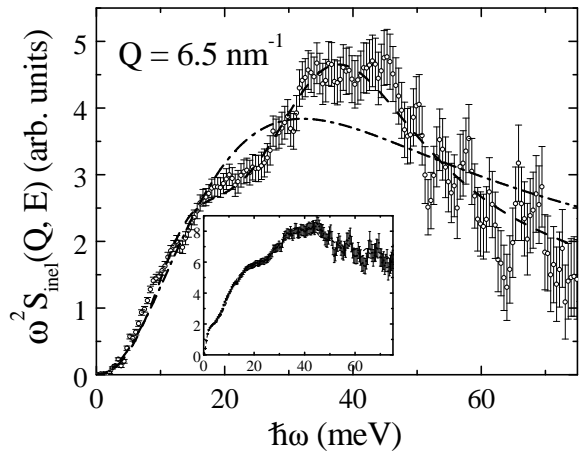}
\vspace{-5.1cm} \caption{The current spectrum of the inelastic
part of the IXS signal (the elastic contribution to the fit is
subtracted by the experimental data) at $Q$=6.5 nm$^{-1}$ (open
circles) is reported together with its $\pm$$\sigma$ error bars
and the one- (dot-dashed) and two- (dashed) excitations fits. The
inset shows the current spectra of the whole IXS signal, without
subtraction of the elastic contribution, and the total fit to the
spectrum (dashed line) including the elastic and inelastic (two-
excitations) part.} \label{fig2}
\end{figure}

All the measured IXS spectra have been fitted using both one and
two inelastic features, and the F-test has been applied to the
results in order to establish whether the second excitation is
statistically necessary to describe the spectra: this is always
the case for $Q$ values larger than 4 nm$^{-1}$. The statistical
significance of the two excitations fit of the spectrum at $Q$=4
nm$^{-1}$ is at the threshold. Previous IXS spectra at lower
$Q$'s \cite{IXSsilica2} have been re-analyzed and no evidence for
the second peak has been found.

It is worth to note that, due to the weakness of the inelastic
signal with respect to the intense elastic peak in this sample,
the difference between the one- and two-excitations fits could
not be appreciated from Fig.~1. However, beyond the quantitative
statistical data analysis, the presence of two excitations clearly
arise looking at the {\it longitudinal current} spectra,
$C_L(Q,\omega)=(\omega^2/Q^2)S(Q,\omega)$, as it is evident in
Fig.~2. As an example, in this figure we report for $Q$=6.5
nm$^{-1}$ the difference, multiplied by $\omega^2$, between the
IXS data and the $S_{el}(Q,\omega)$ determined by the fit (open
circles with error bars). The broken lines represent $\omega^2
S_{inel}(Q,\omega)$ obtained by the previously described fit of
the measured spectra with one (dash-dotted) or two (dashed)
excitations. As can be easily seen, one single excitation does
not account even qualitatively for the measured spectrum. Even if
the agreement between the data and the two excitations fit is not
perfect, pointing out the need for a more complete generalized
hydrodynamic model, it is clear that a two-excitations
description of the spectra is compulsory in this $Q$ region.

To show that the need for a two excitation representation of the
data -as demonstrate in the main panel of Fig.~2- is not an
artifact due to the subtraction of the elastic line, the
"experimental" current spectra (raw IXS data multiplied by
$\omega^2$) is reported in the inset of Fig.~2. The presence of
the (almost Lorentzian) resolution function manifests itself as a
third bump at low energies ($\approx$1.5 meV), and as a long
(almost $\omega$-independent) tail. Beside these two resolution
features, the presence of two excitations is evident also in the
inset of Fig.~2, i.~e. in the raw data.

\begin{figure}[h]
\centering
\includegraphics[width=.43\textwidth]{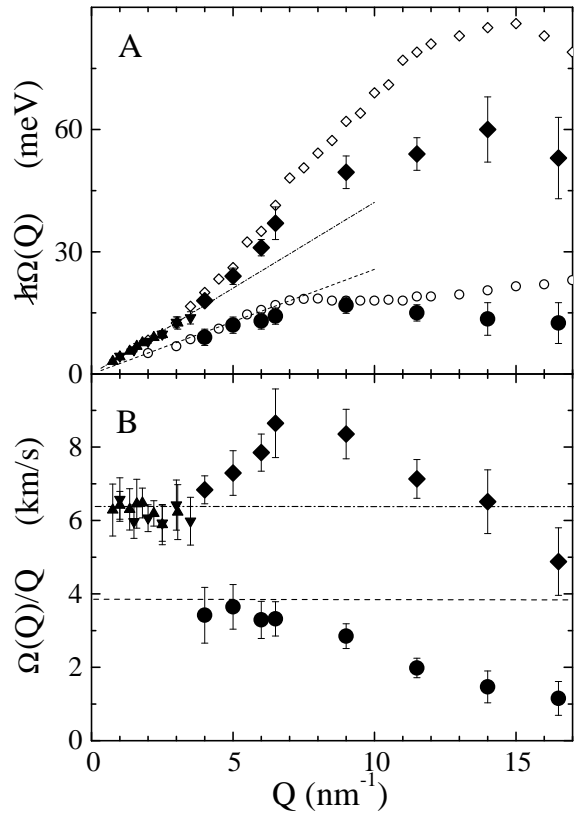}
\caption{A) Dispersion relation of vitreous silica in the first
p-BZ. Full symbols are experimental data (maxima of the two
inelastic contribution to the current spectra) representing the
excitation energies ($\hbar\Omega(Q)$) of the longitudinal
(diamond) and transverse (circle) branches. Data from $Q$=4
nm$^{-1}$ are from present work, below $Q$=4 nm$^{-1}$ are from
Ref.s~\cite{IXSsilica1,IXSsilica2,IXSsilica3} (triangles). The
open symbols are the main maxima of the simulated L (open diamond)
and T (open circle) current spectra from Ref.~\cite{MDpilla}.
Dash-dotted and dashed lines are the extrapolation of the low
frequency L and T sound velocity, respectively. B) Generalized
sound velocity derived from the data of panel A as
$v(Q)$=$\Omega(Q)/Q$. Symbols as in A, the MD data are not
reported.} \label{fig3}
\end{figure}

The excitation energies obtained by the fit and the corresponding
apparent sound velocities ($v(Q)$=$\Omega(Q)/Q$) are reported in
Fig.s~3A and 3B respectively. Data from previous IXS measurements
in vitreous-silica \cite{IXSsilica1,IXSsilica2,IXSsilica3},
performed at low $Q$ are also reported. The corresponding
dispersion curves obtained from numerical studies \cite{MDpilla}
as the maxima of the longitudinal and transverse currents are
plotted in the same figure (open symbols). It is evident from
Fig.~3 that the two excitations found by the fit analysis of the
experimental data match with the LA and TA branches derived from
the MD simulations. In particular, the lower energy excitation in
the IXS data is in very good agreement with the TA branch, and is
then associated with the spilling of the TA excitation in the
measured (longitudinal) spectra due to polarization mixing
\cite{Mixing} that, in this glass, becomes effective for $Q$'s
larger than $\approx$5 nm$^{-1}$.


The excitation at higher energy obtained from the fit of the
measured spectra is associated to the LA mode: the $\Omega(Q)$
values obtained here are the extension at high $Q$'s of the LA
dispersion relation previously determined by IXS
\cite{IXSsilica1,IXSsilica2,IXSsilica3}, and are also in agreement
with the trend of the LA branch derived from MD simulations. This
branch shows a clear bend-up at $\approx$5 nm$^{-1}$, i.~e. the
generalized sound velocity undergoes a speed up from its low $Q$
value ($v$$\approx$6400 m/s) to $\approx$9000 m/s \cite{nota}. The
existence of a positive dispersion of the generalized sound
velocity could be interpreted as the signature of an underlying
relaxation process. This possibility, as well as the origin of
this process -that cannot be identified with the structural
($\alpha$) relaxation process, basically frozen in the glassy
phase- is an open problem. In Ref.~\cite{relHarm} it has been
suggested that the origin of this relaxation process could be
retrieved in the nature of the spatial pattern of the vibrational
eigenmodes of a disordered structure, rather than in dynamical
effects as, for example, the anharmonicity of the vibrational
dynamics. Further study, specifically on the temperature
dependence of this phenomena, are required to asses this point. It
is important to mention that, by a INS study of the dynamics of
$v$-silica at large momentum transfer ($Q>$20 nm$^{-1}$), Arai et
al. \cite{Arai} deduced a generalized sound velocity of
$\approx$9400 m/s, a value compatible with the one found here in
the high frequency side of the transition. Finally, at $Q$ larger
than $\approx$7-8 nm$^{-1}$, the LA dispersion relation shows the
features of the usual acoustic phonons behavior: it starts to bend
down and reaches a maximum at half of the p-BZ edge ($\approx$11
nm$^{-1}$ in vitreous silica).


In conclusion, in the present work we have studied the dynamics in
the intermediate $Q$ region (from 20\% up to the first p-BZ edge)
in vitreous silica. We have shown that, beside the usual
propagating longitudinal sound modes already observed in glasses
at "small" $Q$'s (up to 20\% of the p-BZ edge in $v$-silica), a
second excitation becomes more and more evident in the spectra as
$Q$ is increased. This second excitation -in the $Q$ range where
it is visible- is almost non-$Q$ dispersing and is assigned to the
Transverse Acoustic (TA) branch. The possibility to observe
simultaneously the TA and LA excitations has allowed the detailed
determination of the shape of the LA branch, which shows three
distinct regions:
%
%
i) At $Q$ below 4 nm$^{-1}$ it linearly disperses with a sound
velocity $v$$\approx$6400 m/s, consistently with light scattering
measurements \cite{BLSsilica}. ii) with increasing $Q$, it does
not show tendency to saturation \cite{Courtens_LA=BP}, but, on the
contrary, it reveals a speed-up, and it reaches a generalized
sound velocity of $\approx$9000 m/s. iii) At $Q$ higher than 7
nm$^{-1}$, it starts to follow a crystal-like behaviour with a
slow down and a maximum around half of the p-BZ. The observation
in the LA branch of vitreous silica of a positive dispersion of
the generalized sound velocity provides an important missing tile
in the phenomenology of disordered induced features in glasses. As
a speculation, a possible explanation for such a positive
dispersion can be searched in the presence of a very fast
"relaxation" process associated to the structural disorder, as
suggested in Ref.~\cite{relHarm}, or in the interaction of the LA
branch with other modes, a phenomenon similar to the symmetry
avoided crossing in crystals.


\begin{references}
\bibitem{nota}
It is worth to stress that while the actual values of $\Omega(Q)$
plotted in fig.~\ref{fig3} -and consequently those of $v(Q)$- can
depend on the specific model used to represent the $S(Q,\omega)$,
the existence of a positive dispersion of $v(Q)$ is a robust
result. Indeed, in order to invalidate the existence of a positive
dispersion of the $v(Q)$, the maximum of the spectrum at $Q$=6.5
nm$^{-1}$ (which is found from the fit to be at
$\Omega$$\approx$37 meV) should sit at $\Omega$$\approx$27.5 meV
or less. Looking at the $Q$=6.5 nm$^{-1}$ current spectrum
reported in Fig.~2, one can easily recognize that this not
compatible with the experimental data.
\end{references}
\end{document}